\documentclass[twocolumn,groupedaddress,showpacs,floatfix,notitlepage]{revtex4-1}
\usepackage{graphicx}
\usepackage{epsfig}
\usepackage{amsfonts}
\usepackage{ulem}
\usepackage{amsmath}
\usepackage{subfig}
\usepackage[autostyle]{csquotes}

\usepackage[retainorgcmds]{IEEEtrantools} 
\newcommand{\ud}{\,\mathrm{d}}

\begin{document}

\title{Contact Time Periods in Immunological Synapse}

\author{Daniel R Bush}
\affiliation{                    
Non-linearity and Complexity Research Group - Aston University, Aston Triangle, Birmingham, B4 7ET, UK}
\author{Amit K Chattopadhyay}
\affiliation{                    
Non-linearity and Complexity Research Group - Aston University, Aston Triangle, Birmingham, B4 7ET, UK}
\email{a.k.chattopadhyay@aston.ac.uk}
\begin{abstract}
The letter resolves the long standing debate as to the proper time scale ($<\tau>$) of the onset of the immunological synapse (IS) bond, the non-covalent chemical bond defining the immune pathways involving T-cells and antigen presenting cells (APC).
Results from our model calculations show $<\tau>$ to be of the order of seconds instead of minutes.
Close to the linearly stable regime, we show that in between the two critical spatial thresholds defined by the integrin:ligand pair ($\Delta_2\sim$ 40-45 nm) and the T cell receptor (TCR):pMHC bond ($\Delta_1\sim$ 14-15 nm), $<\tau>$ grows monotonically with increasing co-receptor bond length separation $\delta$ (= $\Delta_2-\Delta_1\sim$ 26-30 nm) while $<\tau>$ decays with $\Delta_1$ for fixed $\Delta_2$. 
The non-universal $\delta$-dependent power-law structure of the probability density function (PDF) further explains why only the TCR:pMHC bond is a likely candidate to form a stable synapse. 
\end{abstract}
\date{\today}


\maketitle

\section{Introduction}
Cell to cell contacts define key chemical pathways that articulate immune response signaling through cellular signal transduction~\cite{bib:janeways_2006}. 
Signals are transported in the event of attached integrin-ligand pairs, whereupon they are carried through intracellular signaling pathways.
Such bio-mechanical signaling, mediated by surrounding coreceptor molecules (CD4, CD8, CD45) proliferates an immune response in the body through cellular level interactions \cite{bib:choudhuri_2014}, often resulting in the formation of the immunological synapse (IS) bond between the immune cells (T \& B cells) and the antigen presenting cells (APC).
In T-cells, some of these signaling pathways are directed toward the nucleus, where conformal changes lead to cell proliferation and triggering of the immune effector functions.
	
The T-cell:APC bond, a \enquote{close contact} patch between the membranes, is known to have a diameter $\sim$10 $\mu$m and contains a large number of membrane bound molecules~\cite{bib:grakoui_1999}.
Important cell surface molecules on the T-cell include the T-cell receptor (TCR) and the leukocyte function associated-1 adhesion molecule (LFA-1).
The T-cell receptor (TCR) binds with peptide bound major histocompatibility complexes (pMHC) on the APC.
Similarly, the integrin LFA-1 molecule has a natural ligand in the intercellular adhesion molecule-1 (ICAM-1) on the APC surface.

Nascent TCR:pMHC and LFA-1:ICAM-1 bonds begin to form following the initial cellular attachment.
Fluorescent tagging reveals heterogeneous segregation and aggregation of the surface molecules throughout the interface \cite{bib:monks_1998,bib:grakoui_1999}, a result that is attributed to the presence of multiple length scales ($\sim$14-15 nm for TCR:pMHC and $\sim$40-45 nm for LFA-1:ICAM-1 \cite{bib:barclay_1997,bib:springer_1990}) leading to the dynamic reorganization of the cell surface molecules as a pathway towards IS formation \cite{bib:bunnell_2010,bib:yokosuka_2010,bib:davis_1996b}.
Initially, the longer bonds (LFA-1:ICAM-1) localize at the centre of the contact zone, with small patches of shorter bonds (TCR:pMHC) toward the edge of the contact zone. Such nonlinear patterning have been studied in details \cite{bib:qi_2001,bib:burroughs_2002} with theoretical models convincingly proving that the dynamical reorganization is the result of the presence of multiple length scales in the problem along with descriptions of conditions in which such changes in receptor configurations \cite{bib:grakoui_1999} occur. In what follows, we will use this fact as an input but otherwise focus on a different aspect of quantifying the strength of the immunological synapse bond that was beyond the scope of either of these papers.

From the binding assay measurements, we know that the affinity of the TCR-pMHC complex may vary drastically depending on the proliferated peptide, TCR sequences and the MHC allele. What triggers the sequential patterning process leading primarily to the aggregation-segregation mechanism, followed by binding of the IS bond, are defined by the binding energies of the respective of the emergent synapse. While such a binding affinity is not exactly proportional to the separation length of the individual cell membranes, it has now been proved \cite{bib:stone_2009} that the TCR:pMHC interaction has to be long enough to complete proximal signalling while the dissociation rates have to be sufficiently short to allow multiple TCRs engaging with the same pMHC. As explained in \cite{bib:stone_2009}, following an initial clustering of the TCRs on the cell surface, affinity rates are affected by the average \enquote{dwell times} with affinity increasing monotonically as dwell times up to a threshold followed by a saturation regime; so while it would be ideal to model the TCR:pMHC interaction with affinity as the switch, defining the average dwell time as our regulation \enquote{order parameter} allows for more direct theoretical modelling of the IS dynamics and in turn the patterning process. At this minimalist level, we are considering only one-no-one TCR-pMHC interactions which is why TCR intensity resulting from the aggregation on the cell membranes will only linearly affect the IS kinetics. 

The early patterning signals cause polarization of the microtubule organizing centre, effectively stopping T-cell migration and orienting the internal machinery toward the contact area.
At a time scale of up to 5 minutes, the pattern inverts such that the TCR:pMHC bonds migrate to the centre of the initial \enquote{close contact} patch and the LFA-1:ICAM-1 bonds form a tight adhesion ring around the periphery \cite{bib:grakoui_1999}. 
Such a {\it pattern inversion} is strictly a \enquote{non-universal} feature, since dynamic interactions (kinapses) between the cells are also shown to initiate immune effector functions~\cite{bib:gunzer_2000,bib:dustin_2008}.

It has been largely recognized that the mature synapse is required to trigger the immune effector function \cite{bib:davis_1996b}.
A primary objective of this study is to focus on the kinetic behavior at the start of this dynamical bond formation process. When the TCR is attached to an agonist pMHC, intracellular signaling molecules can phosphorylate the cytoplasmic portion of the TCR and signal transduction occurs. However, it is not well understood if signaling continues after the TCR has disengaged from the pMHC, although some studies indicate this may be possible ~\cite{bib:bunnell_2010}.
Recent studies have also identified TCR microclusters, small patches of membrane enriched in TCR and signaling molecules, continuously forming at the periphery of the synapse throughout the contact duration\cite{bib:yokosuka_2005}.
Signaling from these microclusters peaks while in the periphery and diminishes as they migrate toward the centre of the synapse.
The time dynamics of these signals are in the order of seconds, consistent with upregulation of Ca$^{2+}$ levels, before the synapse matures minutes later ~\cite{bib:dustin_2008}.

The strength and start time of the IS bond patterning is defined by the average time the two randomly forced (due to thermal fluctuations) fluctuating membranes (T cell \& APC) remain in contact with each other above a minimum threshold $\Delta$ that is defined by the bond lengths of the participating molecules, a model that was successfully implemented previously \cite{bib:chattopadhyay_2007} in estimating the average lengths scale of the interacting T Cell:pMHC patch sizes. 
In line with the Chattopadhyay-Burroughs' model \cite{bib:chattopadhyay_2007}, here we analyze the average time of contact of these \enquote{close contact} patches, at the start of patterning, based on an analogous one membrane-two threshold model (theoretical architecture follows \cite{bib:chattopadhyay_2007}). This is a linearized (around the linearly stable fixed point) version of the nonlinear reaction-diffusion model due to Qi, et al \cite{bib:qi_2001,bib:raychaudhuri_2003,bib:burroughs_2011,bib:burroughs_2002} in line with our previous work in this sequel \cite{bib:chattopadhyay_2007}. Numerical analysis of the nonlinear models \cite{bib:qi_2001,bib:burroughs_2011} have been shown to be in near quantitative agreement with the early images of the synapse\cite{bib:grakoui_1999,bib:monks_1998}.
Our linearized model portrays the non-stationary state dynamics of a fluctuating membrane $\phi(\mathbf{x},t)$ close to the linearly stable point and across a range of mean separation distances defined by the bond lengths of relevant coreceptor molecules (15-45nm).
As shown in \cite{bib:Kardar,bib:qi_2001}, incorporation of the first nonlinear (cubic) perturbation in the linearized (stochastic) model predicts a Hopf-bifurcation point below which the linear regime dominates and above which nonlinear patterning \cite{bib:burroughs_2011} takes over. Our focus here is to study the crossover from the linear to the nonlinear regime. 

\section{The TCR:APC membrane fluctuation model}

\noindent
The dynamics that we are studying here stems from the interaction of two stochastically driven membranes. With a membrane separation distance designated by $\phi({\bf x})$ where ${\bf x}$ is any point on the membrane observed at time $t$, our linearized stochastic continuum model can be written as:
\begin{equation}
M \dot{\phi} = -B \nabla^4 \phi + \gamma \nabla^2 \phi - \lambda \phi + \eta(\mathbf{x},t).
\label{eq:linear_model}
\end{equation}

\noindent
Here $B$ is the coefficient of the membrane rigidity, $\gamma$ is the surface tension, $\lambda$ quantifies the linearized relaxation kinetics close to equilibrium and $M$ is the membrane damping constant. As in a standard membrane dynamics, the membrane rigidity term and the surface relaxation terms create a force balance by working against each other while the contribution from the surrounding coreceptor molecules is encapsulated in the linear $-\lambda \phi$ term.
The thermal noise $\eta(\mathbf{x},t)$ is assumed to be a spatio-temporally independent Gaussian white noise defined through fluctuation-dissipation kinetics \cite{bib:chattopadhyay_2007}
\begin{IEEEeqnarray}{rCl}
\left< \eta \left( \mathbf{x},t \right) \right>  & = & 0
 \IEEEyessubnumber \label{noise_1} \\
\left< \eta \left( \mathbf{x},t \right) ~ \eta \left( \mathbf{x}',t'
 \right) \right>  & = & 2{k}_B{T}~M \delta \left( \mathbf{x} - \mathbf{x}' \right) \delta \left( t-t' \right)
 \IEEEyessubnumber \label{noise_2} \qquad
\end{IEEEeqnarray}

Although broadly argued from a thermodynamic perspective, the above stochastic (model) has its origin in a more detailed nonlinear dynamical system architecture as propounded in \cite{bib:qi_2001,bib:burroughs_2002}. The linearized model focuses on the contact events (TCR:APC synapse) and arises from a linear stability analysis of the nonlinear model defined in these references close to the equilibrium point. Detailed descriptions of the linear stability criteria and its application are available in \cite{bib:burroughs_2002} and have been previously employed in \cite{bib:chattopadhyay_2007}. It must be mentioned that the range of validity of this linear model is limited to the start of the immunological synapse patterning and can not account for the eventual self-organised criticality leading to the mature synapse formation.

\begin{figure}[htp]
	\centering
	\includegraphics[width=0.45\textwidth]{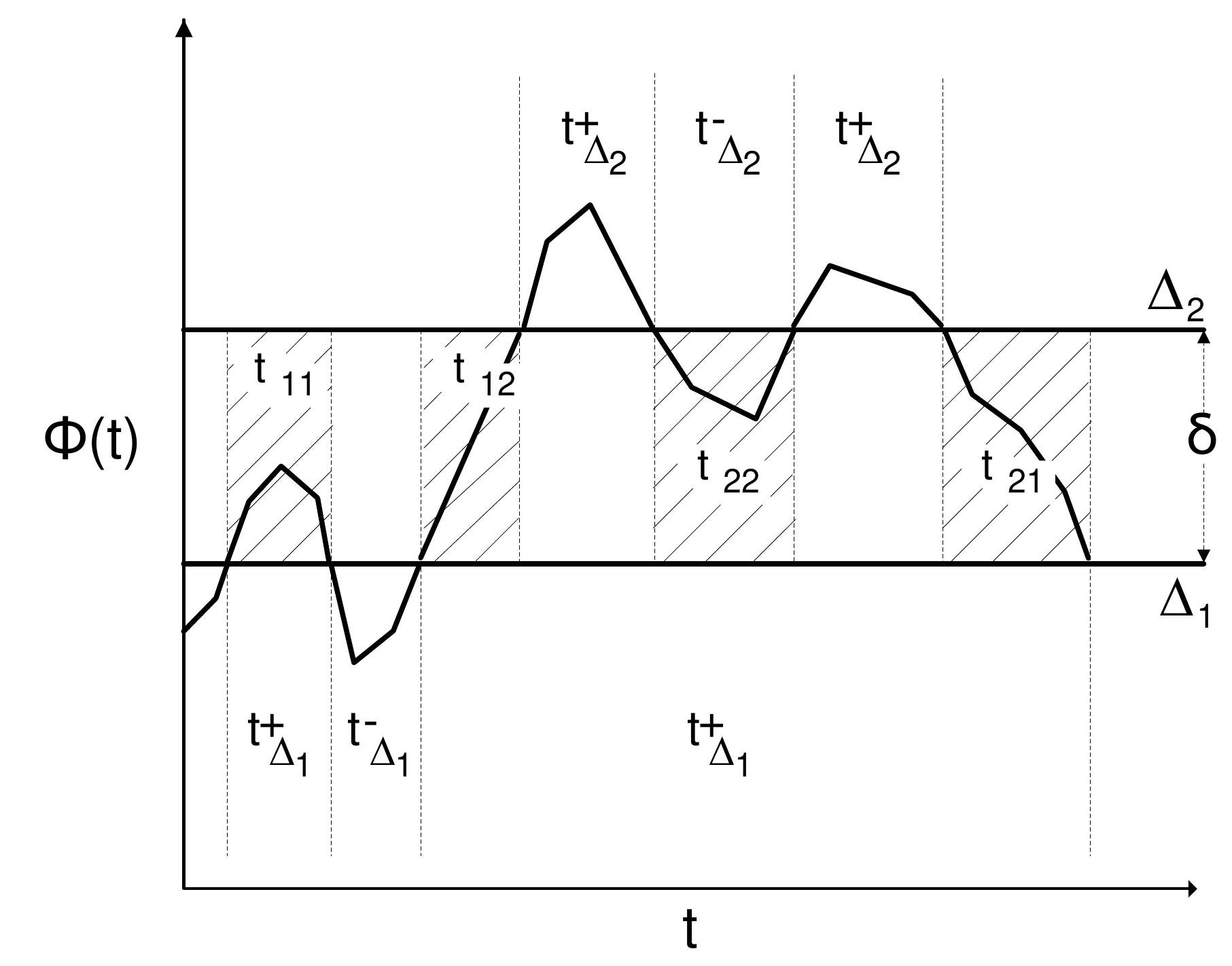}
	\caption{$t^+_{\Delta}$ and $t^-_{\Delta}$ regions: The time evolution of the separation distance for a point in the membrane interface, $\phi(\mathbf{x},t)$.  The $t^+_{\Delta_i}$ regions are periods of time where \enquote{close contact} exists between the membrane surfaces at distances $\Delta_i$ ($i=1,2$) and during the $t^-_{\Delta_i}$ period the membrane separation distance is not favorable for ligand-receptor bond formation.
The shaded regions indicate the time persistence \cite{bib:sire_2004} between two thresholds, $\Delta_1$ and $\Delta_2$.
	\label{fig:t_regions1}}
\end{figure}

\subsection{The two threshold model}

Our starting model is a generalized version of the model described in \cite{bib:chattopadhyay_2007}. In this earlier work, we defined a one membrane model fluctuating across a threshold as an analogue of our physical system. In that model, this single threshold cloned the TCR:APC and other small coreceptor bond length range but considered all larger length bonds (e.g. integrin-ligand) as a constant. This present improved model considers two thresholds $\Delta_1$ (TCR:pMHC) and $\Delta_2$ (integrin:ligand) in acknowledgment of the presence of two different length scales, one large and other small, in a reminder of \cite{bib:burroughs_2011}. 
Two opposing points on the membranes are said to be within \enquote{close contact} if the separation distance is less than $-\Delta_i$ nm (i=1,2), that, in the one membrane model, translates to a configuration of a fluctuating membrane staying above a critical threshold $-\Delta_i$ nm through a certain average distance $<X_{+}>$ \cite{bib:chattopadhyay_2007} and an average time $<\tau^{+}>$ (result from this article).
Fig \ref{fig:t_regions1} explains the time persistence behavior of the fluctuating membrane where ${t^{\pm}}_{\Delta_i}$ (i=1,2) gives the \enquote{bottom-up/up-bottom} cross-over times across the lines $\phi=\Delta_i$, where the duration of time for \enquote{close contact} is given by $t^{-}_{-\Delta_i}=t_2-t_1$.
In accordance with the reflection symmetry of this model, $t^+_{\Delta_i}=t_2-t_1$ represents the \enquote{close contact} duration under the conditions
\begin{IEEEeqnarray}{rCl}
\phi(t_1) = \phi(t_2)=\Delta_i \IEEEyessubnumber
\label{tp_plus_1} \\
\phi(t) \geq \Delta_i, \qquad ~t_1 < t < t_2 \IEEEyessubnumber
\label{tp_plus_2}
\end{IEEEeqnarray}
We shall use the latter definition for notational purposes, as detailed in the figure~\ref{fig:t_regions1} caption.
\subsection{IIA for a two threshold model}
As in \cite{bib:chattopadhyay_2007}, we assume each successive crossing of the $\phi=\Delta_i$ line to be statistically independent ({\it{independent interval approximation}} (IIA) \cite{bib:majumdar_1998,bib:derrida_1996}).
The time persistence characteristics between two thresholds $\Delta_1$ and $\Delta_2$ ($\Delta_2 > \Delta_1$) that are separated by a distance $\delta = \Delta_2 - \Delta_1$ represents a set of
four different events where a signal, $\phi$, persists between the two thresholds (figure~\ref{fig:t_regions1}): 
a) the fluctuation enters the $\delta$ region from below the lower threshold, persists within the $\delta$ region and returns below the lower threshold, $t_{11}$; 
b) the fluctuation enters the $\delta$ region from below the lower threshold, persists and becomes larger than the upper threshold, $t_{12}$; 
c) the fluctuation enters the $\delta$ region from above the upper threshold, persists and exits below the lower threshold, $t_{21}$; 
and finally, d) the fluctuation enters from above the upper threshold, persists and exits above the upper threshold, $t_{22}$.
A consummate representation of the average time of persistence considering all four scenarios together can then be defined as 
\begin{equation}
< t_\delta>=\sum\limits_{i,j=1}^2 w_{ij}(\delta) <t_{ij}(\delta)>,
\label{t_delta}
\end{equation}
where $w_{ij}(\delta)$ is the $\delta$-dependent probability of occurrence of the event $t_{ij}$, with $\sum_{i,j=1}^{2} \:w_{ij}(\delta) = 1$. Detailed quantitative depictions of statistics in each of these zones will be detailed in the following section. In the numerical simulation of the model, the above normalization condition was always adhered to.

\section{Analysis}

The theoretical routine focuses both on the analytical as well as on the numerical aspects. For the former, the target is to recast the model solutions within a Gaussian tuitionary Process (GSP) framework. For the latter, Euler-integration of the stochastic model is followed by the estimation of the probably density function of the persistent crossings. The following sub-sections detail these separately.

\subsection{On to GSP}
In the Fourier transformed space, equation~\eqref{eq:linear_model} admits of a stochastic solution
{\footnotesize
\begin{equation}
\phi \left( \mathbf{x},t \right) =  \frac{1}{2\pi M} \int \ud\mathbf{k} \int_0^t ~ e^{-\alpha \left( \mathbf{k} \right) (t-t') + {i\mathbf{k}\cdot \mathbf{x}}}  \tilde{\eta} \left( \mathbf{k},t' \right) \ud t' + \tilde{\phi}_0
\end{equation}
}
where $\alpha \left( \mathbf{k} \right) = \frac{B \mathbf{k}^4 + \gamma \mathbf{k}^2 + \lambda}{M}$ under initial conditions $\tilde{\phi}_0$. $\phi(\mathbf{x},t)$ is a {\it Gaussian process} with zero mean, whereupon it can be entirely characterized by the two point correlation function (vide \cite{bib:chattopadhyay_2007} for detailed reference; notations are in line with this parent article)
\begin{IEEEeqnarray}{rCl}
c_{12} \left( t_1,t_2\right)  & = & \left< \phi \left( \mathbf{x},t_1\right) \phi \left( \mathbf{x},t_2\right) \right> \nonumber \\
& = & \frac{k_BT}{{\left( 2 \pi\right)}^2 M}\int \text{d}\mathbf{k}  ~\frac{e^{- \alpha(\mathbf{k})(|t_2-t_1|)}}{\alpha(\mathbf{k})}  \nonumber \\
& & - \frac{	k_BT}{{\left( 2 \pi\right)}^2 M}\int \text{d}\mathbf{k}  ~\frac{e^{- \alpha(\mathbf{k})(t_1 + t_2)}}{\alpha(\mathbf{k})}
\label{eqn:temporal_correlation}
\end{IEEEeqnarray}
where $0 < t_1 < t_2$.
In the large time limit, where any of $t_1$ or $t_2$ is large with $|t_2 - t_1|$ still being finite, the second integral term tends to zero and we have a form that is solely dependent on $\tau = |t_2 - t_1|$, which is a {\it Gaussian stationary process} (GSP), with the following correlation functions:
\begin{IEEEeqnarray}{rCl}
c_{12} \left( \tau \right) & = & \frac{k_BT}{{\left( 2 \pi\right)}^2 M}\int \text{d}\mathbf{k}  ~\frac{e^{- \alpha(\mathbf{k})\tau}}{\alpha(\mathbf{k})} \\
c_{12} \left( 0 \right) & = & c_{11} = \frac{k_BT}{{\left( 2 \pi\right)}^2 M}\int \text{d}\mathbf{k}  ~\frac{1}{\alpha(\mathbf{k})}.
\end{IEEEeqnarray}

The above is the representation of our linearized non-equilibrium model in the Gaussian stationary state limit. 
Since we are only interested in \enquote{close contact} times, we introduce the formalism of a conditional correlator for an arbitrary time $t$ during cell-to-cell contact by using the variable $\sigma=\text{sgn}(\phi({\bf x},t)-\Delta_i)$, that changes sign about $\phi({\bf x},t) = \Delta_i$~\cite{bib:majumdar_1998}.
The conditional correlator $A_+(\phi_1, \phi_2)$ = $<\text{sgn}(\phi({\bf x},t_1)-\Delta_i)\;\text{sgn}(\phi({\bf x},t_2)-\Delta_i)>$ for states where $\phi>\Delta_i$ can be expressed as
{\footnotesize
\begin{IEEEeqnarray}{rCl}
A_+(\phi_1, \phi_2) & = & N  \int^{\infty}_{\Delta_i} \text{d}\phi_1  ~e^{-\frac{\text{det} \mathbf{c}}{2 c_{11}} \phi^2_1} \int^{\infty}_{\Delta_i} \text{d}\phi_2  ~e^{-\frac{c_{11}}{2}(\phi_2 + \frac{c_{12}}{c_{11}}\phi_1 )^2}, \nonumber \\
\label{eqn:app:a_plus_final_version}
\end{IEEEeqnarray}
}
with $\text{det}~ \mathbf{c}=c_{11}^2-c_{12}^2$ and the normalization factor
\begin{equation*}
N =\frac{\sqrt{\text{det} \mathbf{c}}}{\text{acot} \left( \frac{c_{12}}{\sqrt{\text{det} \mathbf{c}}}\right)}.
\end{equation*}
As before, $\phi_j=\phi({\bf x},t_j)$ ($j=1,2$) and the curly bracket \enquote{$<>$} refers to ensemble average over all noise realisations.
In the limit of IIA, the probability that the field $\phi$ has crossed the line $\phi=\Delta_i$ in a small enough time interval $\tau$ will be given by $\frac{\tau}{<\tau^+>}$, leading to the relation $A_+=1-\frac{\tau}{<\tau^+>}$ ~\cite{bib:majumdar_1998,bib:derrida_1996}.
This leads to 
\begin{equation}
<\tau^+>=-\frac{1}{{A'}_+}
\label{eqn:analytical_tplus}
\end{equation}
where ${A'}_+$ is the partial derivative of $A_+$ with respect to $\tau$.
In practice, due to the lack of a closed form solution, ${A'}_+=\frac{\partial A_+}{\partial c_{12}} \cdot \frac{\partial c_{12}}{\partial \tau}$ is being evaluated numerically.

\subsection{Numerical scheme}
Continuum equation~\eqref{eq:linear_model} is discretized and simulated using a first-order forward difference Euler scheme 
\begin{equation}
\phi_{i,j} (t+\Delta t) = \phi_{i,j} (t) + \Delta t ~\dot{\phi}_{i,j} (t)
\label{numerical_scheme}
\end{equation}
on a 2-dimensional mesh with $\Delta x = \Delta y = 1$, using periodic boundary conditions.
We use the experimental parameter settings in accord with cellular membranes~\cite{bib:chattopadhyay_2007}: $M=4.7\times10^6$~{\small k}$_B${\footnotesize T}~$s$~$\mu\text{m}^{-4}$, $B=11.8$~{\small k}$_B${\footnotesize T}, $\gamma=5650$~{\small k}$_B${\footnotesize T}~$\mu\text{m}^{-2}$ and $\lambda=6.0\times10^5$~{\small k}$_B${\footnotesize T}~$\mu\text{m}^{-4}$.
For each term to contribute to the dynamics, simple dimensional analysis will show that $B\nabla^4[\phi] \sim \gamma\nabla^2[\phi] \sim \lambda [\phi]$. The numbers here are all in energy units that help us to non-dimensionalize the eventual outcomes. 
The noise in this case is perturbative whose strength is just enough to stimulate dynamical fluctuations. In what follows, we compare results obtained from the approximate analytical solutions (IIA based) with the numerical evaluation of the starting model equation.

\section{Results}
\begin{figure}[htp]
	\centering	
	\includegraphics[width=0.45\textwidth]{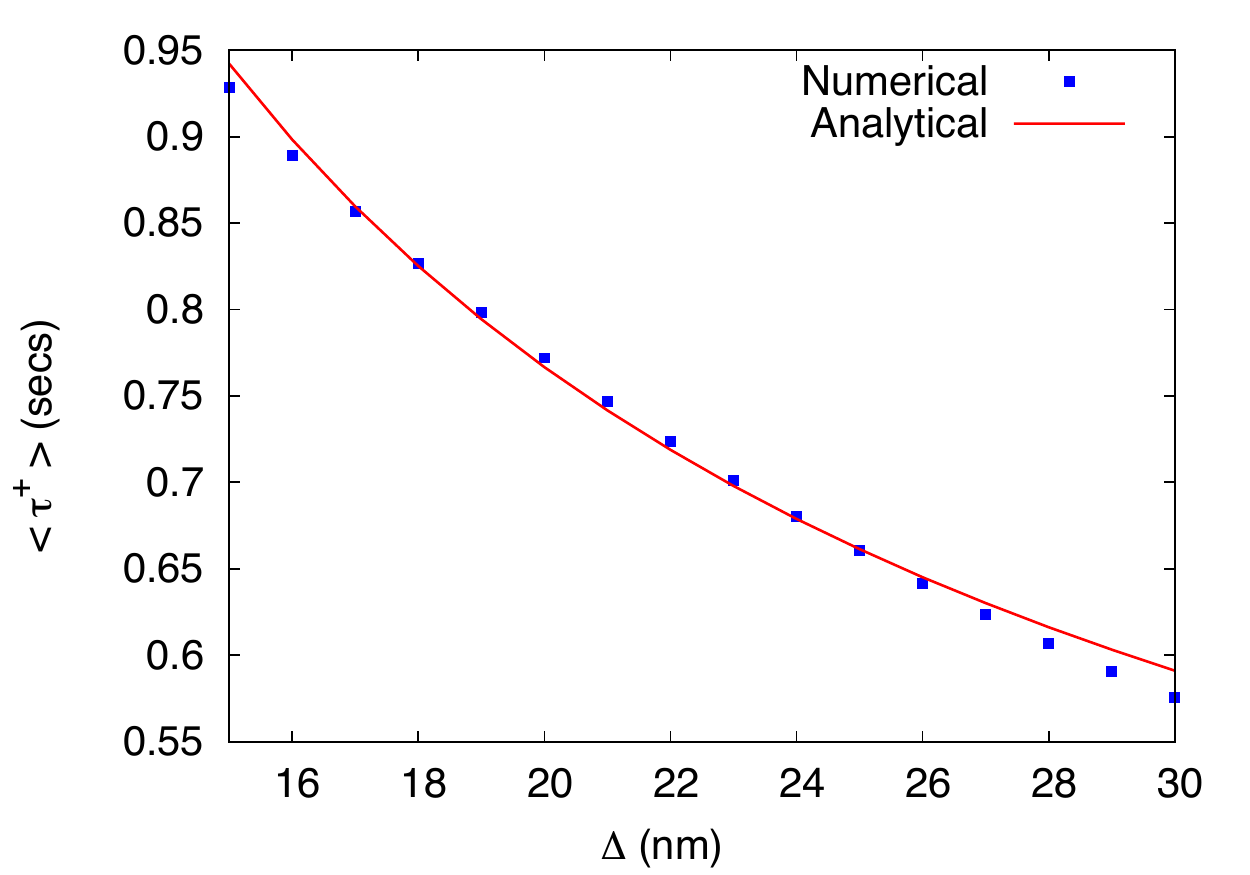}
	\caption{Time persistence for different thresholds,  $<\tau^+>$  vs $\Delta_i$: The ensemble average time persistence above the $\Delta_i$ threshold.
The dots show the result obtained by numerically solving eqn (\ref{eq:linear_model}) based on the scheme detailed in eqn (\ref{numerical_scheme}) while the solid line indicates the analytical result obtained from a solution of eqn (\ref{eqn:analytical_tplus}). The results have been linearly scaled for comparison.
	\label{fig:tplus_vs_delta}}
\end{figure}

Figure~\ref{fig:tplus_vs_delta} shows the ensemble average for the time persistence above any of the ($\Delta_1$ or $\Delta_2$) thresholds keeping the other fixed, for both the numerical and the scaled analytical solution.
The parameters used give rise to persistent \enquote{close contact} patches in the order of magnitude required for the biological problem, that is 10-50 nm. 
The numerical results (in dots: figure \ref{fig:tplus_vs_delta}) are consistent with the analytical solutions (continuous line: figure \ref{fig:tplus_vs_delta}), where the \enquote{close contact} time decreases as the separation distance increases,
suggesting that the TCR:pMHC bond persists longer than the LFA-1:ICAM-1. 

\par
The result shown in figure~\ref{fig:tplus_vs_delta} can be qualitatively understood from simple probabilistic considerations. As the threshold value increases, it becomes more difficult for the randomized (Gaussian) fluctuations to cross this threshold, resulting in reduced average time spent above the threshold value. More non-trivial, though, is the functional nature of the decay in the $<\tau^+>$ value against $\Delta$. As opposed to a simplistic (and incorrect) visual impression, the decay profile here is not exponential, rather it is defined through an intricate balance between power-law scaled fluctuations against the statistics of deterministically decaying membrane fluctuation modes.


A vital part of this model study is the analysis of the dynamics of the randomly driven membrane in between the two given thresholds. In line with the parlance used previously as well as in \cite{bib:chattopadhyay_2007}, this can be represented by an estimation of the time persistence between two threshold values, $\Delta_1$ and $\Delta_2$, where $\Delta_1$ is the T cell analogue of the TCR:pMHC bond length ($\sim$ 15 nm) while $\Delta_2$ symbolizes the ICAM-1:LFA-1 bond separation length ($\sim$ 45 nm).

\begin{figure}[htp]
	\centering
	\includegraphics[width=0.45\textwidth]{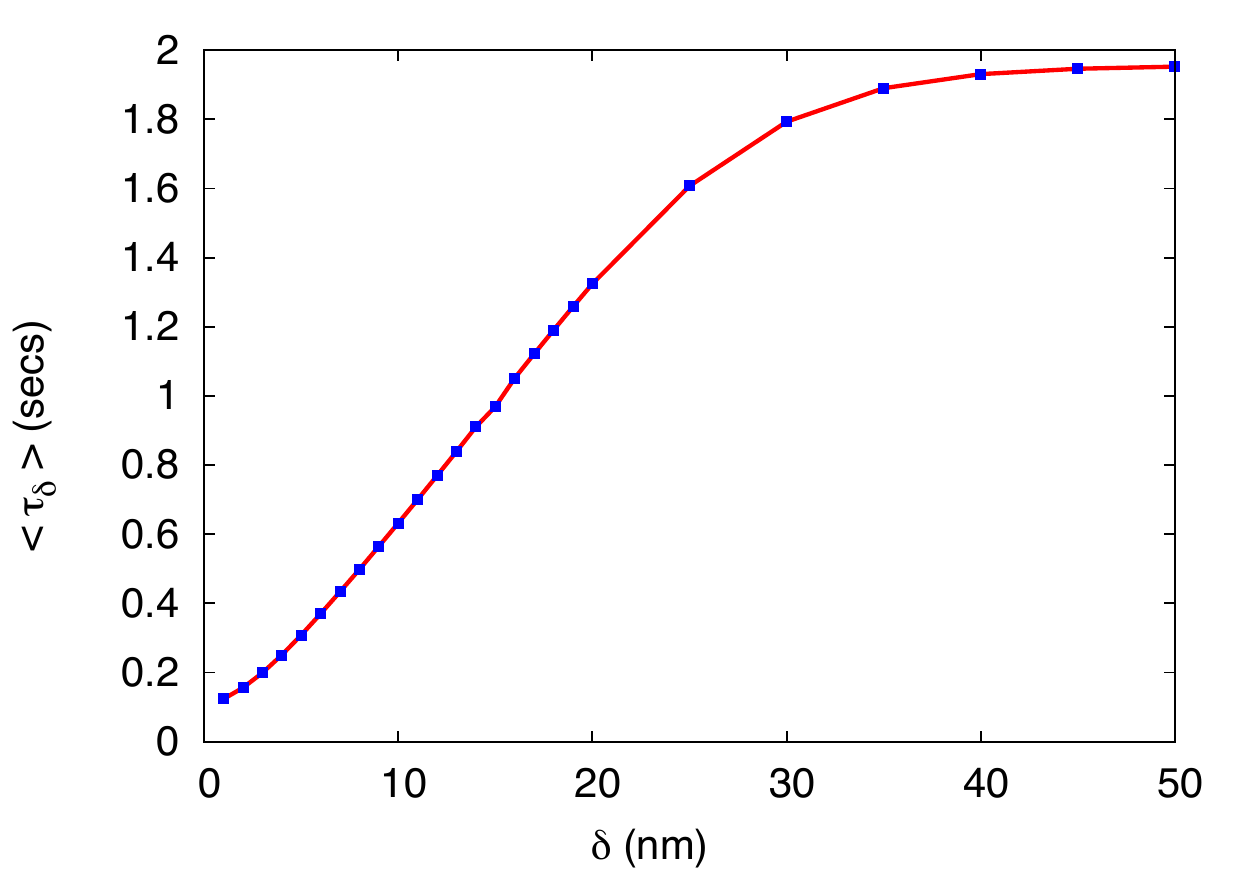}
	\caption[Time persistence, $< \tau_{\delta} >$  vs $\delta$]{Time persistence, $< \tau_\delta >$  vs $\delta$: The ensemble average time persistence between two thresholds, where the distance between the thresholds is given by $\delta$.}
	\label{fig:tb_vs_delta}
\end{figure}

The result for the variation of the average time between the two thresholds as a function of the distance $\delta$ between the thresholds is shown in figure~\ref{fig:tb_vs_delta}. The calculations were done by starting with $\Delta_1=1$ nm and then varying $\Delta_2$ between 1 nm to 50 nm. The results shown in figure~\ref{fig:tplus_vs_delta} are the average over multiple such initial choices of $\Delta_1$ and then varying $\Delta_2$ accordingly to generate the appropriate range for $\delta$.

As previously, the dotted points refer to the numerical simulation results while the continuous line represents the interpolation of the same to maintain continuity.
The time persistence initially increases as $\delta$ increases, then it saturates and asymptotically approaches the $<\tau^+>_{\Delta_1}$ value.
As $\delta \rightarrow 0$ the average time persistence tends toward the smallest time length scale used, for our case $dt=0.1$. Once again, a quantitative understanding of figure~\ref{fig:tb_vs_delta} can be had from the fact that an increase in the $\delta$ value can be wrought about in either of the two possible ways - an increase in $\Delta_2$ for fixed $\Delta_1$, or else a decrease in $\Delta_1$ for a fixed $\Delta_2$. For the first case when $\Delta_2$ increases at a fixed $\Delta_1$, it is easier for a fluctuation mode to remain within the upper limit ($\Delta_2$) than to cross it while for the second case, with a fixed value of the upper threshold $\Delta_2$, a lowering of $\Delta_1$ increases the probability of a fluctuation mode crossing this line resulting in an increased value of $<\tau_\delta>$.

\begin{figure}[htp]
	\centering
	\includegraphics[width=0.45\textwidth]{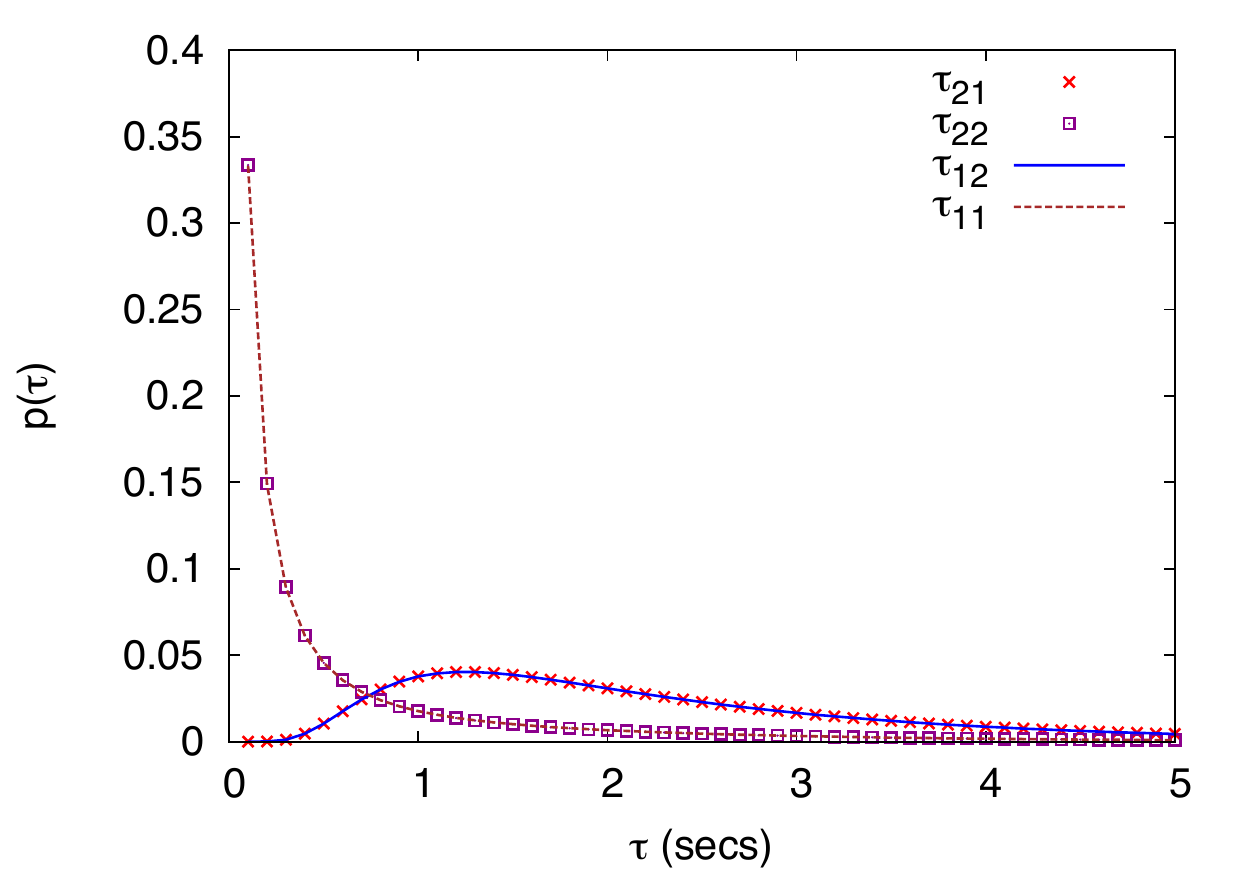}
	\caption{Variation of the probability density function $p(\tau)$ with $\tau$ for $\delta$=15 nm between two thresholds for the $\tau_{11}$, $\tau_{12}$, $\tau_{21}$ and $\tau_{22}$ cases.
	\label{fig:tb_stat_comp}}
\end{figure}

Figures~\ref{fig:tplus_vs_delta} and \ref{fig:tb_vs_delta} respectively express the variation of the average \enquote{persistent times} against bond lengths above a critical threshold and that in between two thresholds ($\Delta_1$ and $\Delta_2$). These results are subsets of a bigger ensemble defined by a membrane that fluctuates across two thresholds $\Delta_1$ and $\Delta_2$, or else that of two membranes whose fluctuations are measured across a single threshold $\Delta$, that, as already explained earlier, are equivalent analytical descriptions. In line with our model of a single membrane fluctuating across two threshold detailed in the previous section and as depicted in figure~\ref{fig:t_regions1}, the statistics can be classified in to four broad zones - \enquote{11}, \enquote{12}, \enquote{21} and \enquote{22}. While \enquote{11} defines the fluctuation regime for a crossing from a region in $\phi<\Delta_1$ across the line $\phi=\Delta_1$ but for $\phi<\Delta_2$, \enquote{22} encapsulates the complementary regime for a crossing across $\phi=\Delta_2$ from a point $\phi>\Delta_2$. \enquote{12} and \enquote{21} represent statistics when crossings are restricted within $\Delta_1<\phi<\Delta_2$ as explained in figure~\ref{fig:t_regions1}. Representing the corresponding average \enquote{persistent time} scales by $\tau_{ij}$ (i,j=1,2), we find that due to reasons of reflection symmetry in the time correlators ($A_{+}(t_+,\Delta_1)=-A_{-}(t_-,\Delta_2)$ and $A_{-}(t_+,\Delta_1)=-A_{+}(t_-,\Delta_2)$), $\tau_{11}$ and $\tau_{22}$ are identical, as are $\tau_{12}$ and $\tau_{21}$. Comparing with the notations used previously, $\tau_{12}(=\tau_{21})$ may be identified with $\tau_\delta$ while $\tau_{11}(=\tau_{22})$ may be identified with $<\tau^+>$ under the constraint $\phi<\Delta_2$. For the same reason (reflection symmetry), as shown in figure~\ref{fig:tb_stat_comp}, the respective density distributions too conform to these symmetry lines. 

\begin{figure}[htp]
	\centering
	\includegraphics[width=0.45\textwidth]{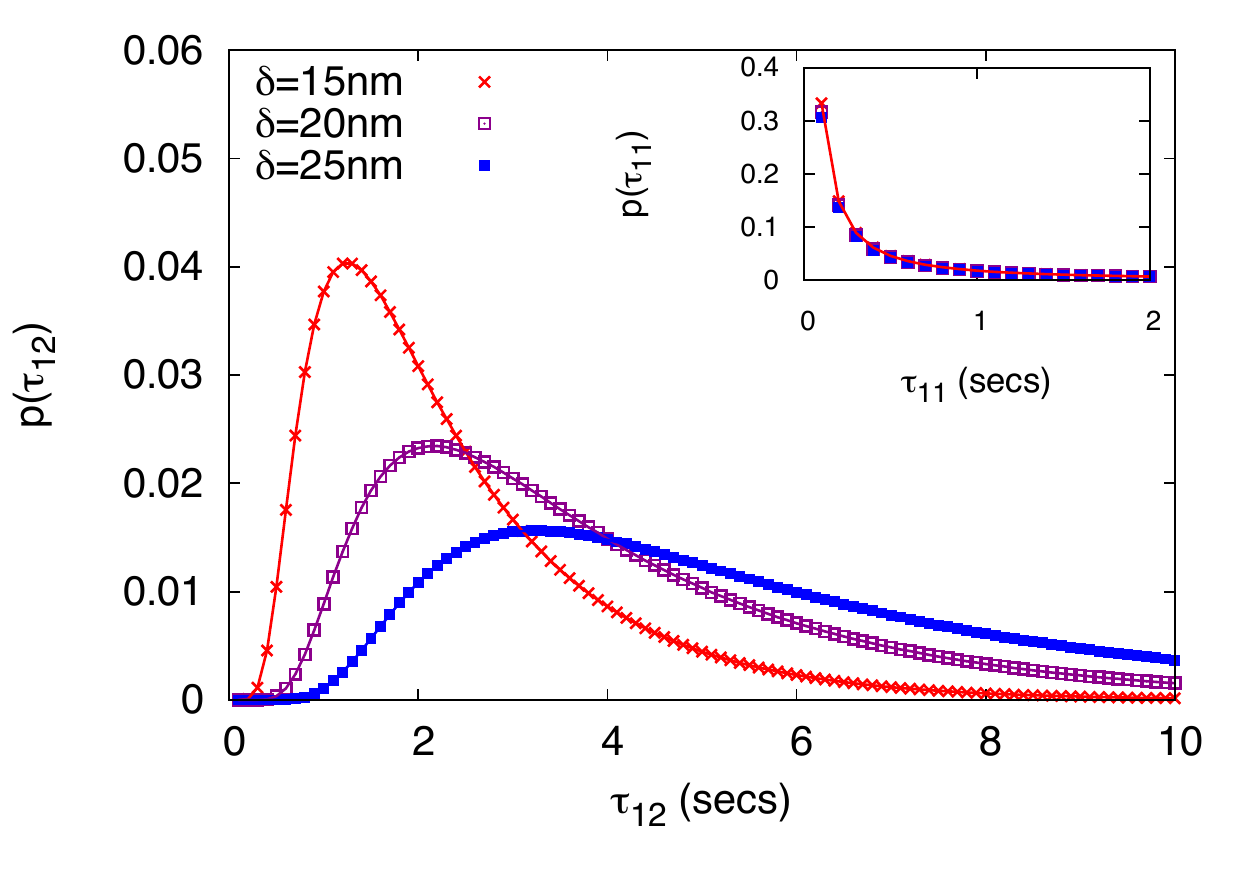}
	\caption{Variation of the probability density function $p(\tau_{12})$ with time $\tau_{12}$ between thresholds $\Delta_1$ and $\Delta_2$ for a range of $\delta$ values. As expected, the smallest $\delta$ (=15 nm) turns up the highest peak (represented by crosses), followed by $\delta$=20 nm (squares) with $\delta$=25 nm showing the lowest peak with the maximum spread (triangles).}
	\label{fig:fig5}
\end{figure}

The result presented in figure~\ref{fig:tb_stat_comp} is not a special case, as shown in figure~\ref{fig:fig5}, where the $\tau_{11}$ and $\tau_{12}$ scenarios are plotted for three values of $\delta$.

As $\delta$ increases the probability density for the $\tau_{12}$ case shows an increase in the probability for longer time persistence, a result that matches with the observation presented earlier through figure~\ref{fig:tb_vs_delta}.
On the other hand, the number of $\tau_{12}$ events  become less frequent as $\delta$ is increased.
The $\tau_{11}$ densities also change slightly, again showing an increase in the probability for longer time persistence.

\section{Conclusions}
The analysis presented here has two major immunological implications. 
Firstly, figures \ref{fig:tplus_vs_delta} and \ref{fig:tb_vs_delta} clearly prove that the onset of patterning at the immature kinapse level, when the central LFA-1:ICAM-1 bond gives way to the smaller TCR:pMHC bond, occurs at the time scale of seconds. 
This discovery is expected to confirm the start time of mature synapse formation. Admittedly, though, the parameter values used make the result a subjective case in that the time scale predicted for a different membrane:membrane dynamics may as well be in minutes or hours, instead of in seconds. Our calculation 
pins down the time scale to within 2-4 seconds that is a further
improvement on the $<$12 seconds' window as suggested in \cite{bib:stone_2009}.
Secondly, the non-universal character of the time correlation of the IS bond as evident in the dependence on the separation length ($\delta$) and separation times ($\tau$), ensuring that the probability density $p(t)$ is a function of the system parameters along with being functions of $\delta$ (figure \ref{fig:tb_vs_delta}) and $\tau_{12}$  (figure \ref{fig:fig5}), confirms a widely acknowledged belief in the community that the TCR:pMHC bond is non-self-organizing in nature and hence is the only stable bond at this spatio-temporal regime.
An immediate impact of this can be seen in the projected time scale for the crossover from the linear to the nonlinear phase that, as is shown in Fig \ref{fig:tb_vs_delta}, spans 2-4 seconds (peak time of the PDF profile). The implication of this analysis is that of a time scale difference of an order of magnitude related to the start time of the \enquote{immature} IS bond formation, a time scale that is also associated with the transition from the linear to the nonlinear regime (and hence our emphasis on a study of the linear stability regime of an otherwise nonlinear dynamics). As to how such crossover is affected by kinase-phosphatase pathways (these pathways act as signal transduction inhibitors and thereby control the rate of the IS bond formation) and what modification this may bring about in the prediction of the time scale of a mature IS bond are some of the exciting topics that we are presently working on. New results are shortly to be communicated on the quantitative nature of the extremal value statistics of these fluctuations and how such non-universal exponents affect the life time and strength of IS bonds.

\section{Acknowledgments}
The authors acknowledge Aston University Research Grant for funding this project. AKC acknowledges insightful discussions with Darren Flower.


\begin{thebibliography}{0}

\bibitem{bib:janeways_2006} Murphy, K., Travers, P. and Walport, M., {\it Janeway's Immunobiology}, published by Garland Science, Taylor \& Francis, New York \& London (2006).

\bibitem{bib:choudhuri_2014} Choudhuri, K., Llodra, J., Roth, E. W., Tsai, J., Gordo, S., Wucherpfennig, K.W., Kam, L. C., Stokes, D. L. and Dustin, M. L., Nature {\bf 507}, 118 (2014).

\bibitem{bib:grakoui_1999} Grakoui, A., Bromley, S. K., Sumen, C., Davis, M. M., Shaw, A. S., Allen, P. M., \& Dustin, M. L.,
{\it Science} {\bf{285}}, 221 (1999).

\bibitem{bib:monks_1998} Monks, C. R. F., Frieberg, B. A., Kupfer, H., Sciaky, N., \& Kupfer, A.,
{\it Nature} {\bf{395}}, 82 (1998).

\bibitem{bib:barclay_1997} Barclay, A. N., Brown, M. H., Law, S. K. A., McKnight, A. J., Tomlinson, M. G., \& van der Merwe, P. A., {\it The Leucocyte Antigen Factsbook. 2nd ed. Factsbook Series.} London: Academic Press Ltd. (1997).

\bibitem{bib:springer_1990} Springer, T. A., {\it Nature} {\bf{346}}, 425 (1997).

\bibitem{bib:bunnell_2010} Bunnell, S. C., {\it Immunological Synapse}, edited by Saito, B., Springer, London (2010).

\bibitem{bib:yokosuka_2010} Yokosuka, T. \& Saito, T., {\it Immunological Synapse}, Eds(Saito, Batista), Springer: London, doi: 10.1007/978-3-642-03858-7\_5 (2010).

\bibitem{bib:davis_1996b} Davis, S. J. \& van der Merwe, P. A., {\it Nat. Immunol.} {\bf{7}(8)}, 803 (2006).

\bibitem{bib:dustin_2008} Dustin, M. L., {\it Immunological Reviews} {\bf{221}}, 77 (2008).

\bibitem{bib:gunzer_2000} Gunzer, M., Sh\"afer, A., Borgmann, S., Grabbe, S., Z\"anker, K. S., Br\"ocker, E.-B., Friedl, P.,
{\it Immunity} {\bf{13}}, 323 (2000).

\bibitem{bib:qi_2001} Qi, S. Y., Groves, J. T., \& Chakraborty, A.,
{\it Proc. Natl. Acad. Sc. (USA)} {\bf{98}}(12), 6548 (2001).

\bibitem{bib:burroughs_2002} Burroughs, N. J. \& W\"ulfing, C., Biophys. J. {\bf 83}, 1784 (2002).

\bibitem{bib:stone_2009} Stone, J. D., Chervin, A. S. \& Kranz, D. M., Immunology {\bf 126}(2), 165 (2009).

\bibitem{bib:yokosuka_2005} Yokosuka, T., Sakata-Sogawa, K., Kobayashi, W., Hiroshima, M., Hashimoto-Tane, A., Tokunaga, M., Dustin, M. L., \& Saito, T.,
{\it Nat. Immunol.} {\bf{6}}(12), 1253 (2005).

\bibitem{bib:chattopadhyay_2007} Chattopadhyay, A. K. \& Burroughs, N. J.,
{\it Europhys. Lett.} {\bf{77}}, 48003 (2007).

\bibitem{bib:Kardar} Ko$\breve{s}$mrlj, A., Kardar, M. \& Chakraborty, A. K., {\it Ann. Rev. Cond. Mat. Phys} {\bf 4}, 339 (2013).

\bibitem{bib:burroughs_2011} Burroughs, N. J., K\"ohler, K., Miloserdov, V., Dustin, M. L., van der Merwe, P. A., \& Davis, D. M.,
{\it PLoS Computational Biology},{\bf{7}}(8), e1002076 (2011).

\bibitem{bib:raychaudhuri_2003} Raychaudhuri, S., Chakraborty, A. K., \& Kardar, M.;
{\it Phys. Rev. Lett.} {\bf{91}}(20), 208101 (2003).

\bibitem{bib:sire_2004} Sire, C., {\it Phys. Rev. Lett.} {\bf 93}, 130602 (2004).

\bibitem{bib:majumdar_1998}  Majumdar, S. N. \& Bray, A. J.,
{\it Phys. Rev. Lett.} {\bf{81}}(13), 2626 (1998).

\bibitem{bib:derrida_1996} Derrida, B., Hakim, V. \& Zeitak, R.,
{\it Phys. Rev. Lett.} {\bf{77}}(14), 2871 (1996); 

\end{thebibliography}
\end{document}